\documentclass[11pt]{cernrep}
\usepackage{graphicx}
\graphicspath{{./pics/},{./fig/}}
\usepackage{isolatin1}
\usepackage{here}
\usepackage{floatflt}
\usepackage{multicol}
\usepackage{amsmath}
\def\NIMA{{\em Nucl. Instrum. Methods} A}

\begin{document}
 \title{DOUBLY CHARMED BARYONS IN COMPASS}
\author{{\em L.~Schmitt, S.~Paul, R.~Kuhn}\\
TU-München, Physik-Department E18, Munich, Germany\\
{\em M.~A.~Moinester}\\
School of Physics and Astronomy, Tel Aviv University, Ramat Aviv, Israel}
\maketitle
\begin{abstract}
The search for doubly charmed baryons has been a topic for COMPASS
from the beginning. Requiring however a complete spectrometer and
highest possible trigger rates this measurement has been postponed.
The scenario for such a measurement in the second phase of COMPASS is
outlined here. First studies of triggering and simulation of the
setup have been performed. New rate estimates based on recent
measurements from SELEX at FNAL are presented.
\end{abstract}

\section{INTRODUCTION}
The COMPASS collaboration was founded in 1996 to perform a number of
measurements in hadron physics ranging from polarized structure
functions examined with deep inelastic muon scattering to topics like
light meson spectroscopy and the study of exotic hadrons \cite{compass}.
After the first phase of the COMPASS experiment focusing on the
contribution of gluons to the polarized structure function of the
nucleon, a second phase is planned to address more topics with hadron
beams. One of these topics, the search for doubly charmed baryons, is
described in this report.\\

The quantum chromodynamics hadron spectrum includes doubly charmed
baryons (DCBs): $\Xi_{cc}^{+}$ (ccd), $\Xi_{cc}^{++} (ccu), $ and
$\Omega_{cc}^{+}$ (ccs), as well as the triply charmed $\Omega_{ccc}^{++}$
(ccc).  A 1996 DCB review \cite{Moines} collected information on masses,
lifetimes, internal structure, production cross sections, decay modes,
branching ratios, yields, and experimental requirements for optimizing the
signal and minimizing the backgrounds. DCB works published since then are
given in Refs.  \cite{kis,ebert,gunter,oni,Gub,Gunt,Itoh, BPhys,nt}.  
The doubly and triply charmed baryons provide a new window for understanding
the structure of all baryons.  As pointed out by Bjorken \cite{bj}, one
should strive to study the triply charmed (ccc) baryon. Its excitation
spectrum, including several narrow levels above the ground state, should be
closer to the perturbative regime than is the case for the DCBs. The (ccq)
studies are a valuable prelude to such (ccc) efforts.

Hadron structures with size scales much less than 1/$\Lambda_{qcd}$
should be well described by perturbative QCD. The tightly bound color
antitriplet (cc)$_{\bar{3}}$ diquark in (ccq)  may satisfy this condition.
But the DCB radius may be large, if it is dominated by the low mass q
orbiting the tightly bound (cc) pair. The study of such configurations and
their weak decays can help to set constraints on models of quark-quark
forces \cite{fr,ros}.  Stong \cite{stong} emphasized how the QQq
excitation spectra can be used to phenomenologically determine the QQ
potential, to complement the approach taken for $Q\bar{Q}$ quarkonium
interactions.

Savage and Wise \cite{sw} discussed the (ccq) excitation spectrum
for the q degree of freedom (with the (cc) in its ground state) via the
analogy to the spectrum of $\bar{Q}q$ mesons, where the (cc) pair plays the
role of the heavy $\bar{Q}$ antiquark.  Fleck and Richard \cite{fr}
calculated excitation spectra and other properties of (ccq) baryons for a
variety of potential and bag models, which describe successfully known
hadrons.  In contrast to heavy mesons, the descriptions of light quark
(qqq) and singly charmed (cqq) baryons are less successful.  We need to
better understand how the proton and other baryons are built from quarks.
The investigation of the (ccq) system should help put constraints on baryon
models, including light quark (qqq) and singly charmed (cqq)  baryons,
since the (ccq) has a quark structure intermediate between (qqq) proton and
$\bar{Q}q$ meson structures.

In the double-charm system, there have been many predictions for
the masses of the J=1/2 states and the J=3/2 hyperfine excitations
\cite{BPhys}.  Most results are consistent with expectations of a ground
state mean mass around 3.6 GeV/$\rm{c}^2$.  The (cc)  color antitriplet
diquark has spin S=1. The spin of the third quark is either parallel
(J=3/2) or anti-parallel (J=1/2) to the diquark.  For (ccq), the J=1/2
states are expected to be lower than the J=3/2 states by around 80
MeV/$\rm{c}^2$ \cite{BPhys,nt,fr,lich}.

Bjorken \cite{bj} and also Fleck and Richard \cite{fr} suggest
that internal W exchange diagrams in the $\Xi_{cc}^{+}$ decay could
reduce its lifetime to around $~100$ fs, roughly half the lifetime of the
$\Lambda_{c}^{+}$.  Considering possible constructive interference
between the W-exchange and two c-quark decay amplitudes, it is possible
that this state should have an even shorter lifetime.

 We describe qualitatively the perturbative production mechanism for DCBs.  
One must produce two c quarks (and associated antiquarks), and they must
join to a tightly bound, small size anti-triplet pair. The pair then joins
a light quark to produce the final (ccq).  The two c-quarks may be produced
(initial state) with a range of separations and relative momenta (up to say
tens of GeV/c). In the final state, if they are tightly bound in a small
size (cc) pair, they should have relative momentum lower than roughly 1
GeV/c. The overlap integral between initial and final states determines the
probability for the (cc)-q fusion process.  Such cross sections may be
smaller by as much as 10$^{-2}$-10$^{-3}$ compared to single-charm
production.  Aoki et al. \cite{aok} reported a low statistics measurement
at $\sqrt{s}$ =26 GeV/$\rm{c}^2$ for the ratio of double to single open
charm pair production, of $10^{-2}$. This $D\bar{D}D\bar{D}$ to $D\bar{D}$
cross section ratio was for all central and diffractive events. This high
ratio is encouraging for (ccq) searches.  Cross section guestimates are
given in Ref. \cite{Moines}.

Brodsky and Vogt \cite{bro} suggested that there may be significant
intrinsic charm (IC) $c\bar{c}$ components in hadron wave functions, and
therefore also $cc\bar{c}\bar{c}$ components.  The double intrinsic charm
component can lead to (ccq) production, as the (cc) pairs pre-exist in the
incident hadron.  Intrinsic charm (ccq) production, with its expected high
X$_f$ distribution, would therefore be especially attractive. When a double
charm IC state is freed in a soft collision, the charm quarks should also
have approximately the same velocity as the valence quark. Thus,
coalescence into a (ccq) state is likely.  Cross section guestimates are
given in Ref. \cite{Moines}.

The semi-leptonic and non-leptonic branching ratios of (ccq) baryons were
estimated by Bjorken \cite{bj} in 1986. He uses a statistical approach to
assign probabilities to different decay modes. He first considers the most
significant particles in a decay, those that carry baryon or strangeness
number. Pions are then added according to a Poisson distribution. The
Bjorken method and other approaches for charm baryon decay modes are
described by Klein \cite{kle}.  For the $\Xi_{cc}^{++}$, Bjorken \cite{bj} 
estimated the $\Lambda_c^{+} \pi^{+} K^{-} \pi^{+}$ final state to
have 5\% branching ratio; while for the $\Xi_{cc}^{+}$, he estimated the
$\Lambda_c^{+} \pi^{+} K^{-}$ final state to have 3\% branching ratio. One
expects \cite{Moines} that roughly 80\% of the (ccq)  decays are hadronic,
with as many as one-third of these leading to final states with all charged
hadrons.

Recently the SELEX experiment at Fermilab has reported the first
observation of the doubly charmed baryon $\Xi_{cc}^+$ in the channel
to  $\Lambda_c K^-\pi^+$ \cite{mattson}.
Evidence for other states was found as well. The forward production
seems to be strongly enhanced in baryon beams. Effectively, a large
fraction of their observed $\Lambda_c$ are daughters of doubly charm
baryons \cite{russ}. This requires new mechanisms of charm- or even 
di-charm generation. Therefore much larger yields of
doubly charmed baryons could be expected at the high rate COMPASS
experiment.

\section{EXPERIMENTAL SETUP}
To measure doubly charmed baryons with the COMPASS spectrometer
several modifications have to be made and some detector systems 
have either to be built from scratch or upgraded. This section
outlines the hardware requirements for a DCB measurement.

It is foreseen to run the double charm measurement with a proton
beam of 280 GeV (the maximum of the present beamline setup) and
an intensity of up to $10^8$ during the 5 s SPS spill every 16.8s.
\begin{figure}[h]
  \begin{center}    
  \includegraphics[width=12cm]{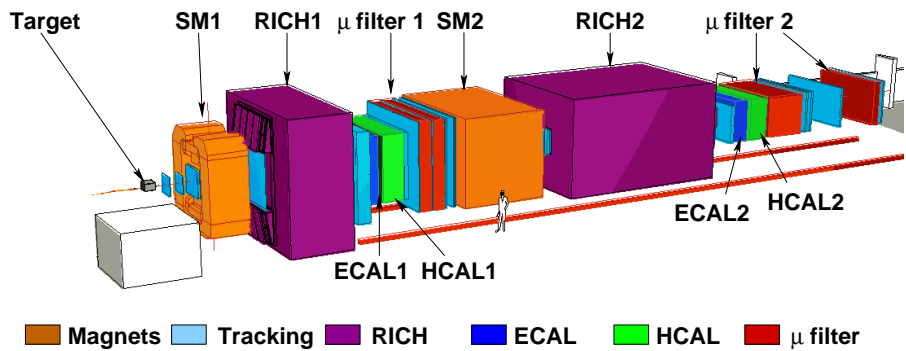}
    \caption{General setup of the COMPASS spectrometer.}
    \label{fig:setup}
  \end{center}
\end{figure}
\subsection{Spectrometer}
% Tracking
% PID
% Trigger
% - ECAL1, HCAL1 = Et
% - MW1, Hodoscope
COMPASS uses a double magnetic spectrometer with tracking,
electromagnetic and hadronic calorimetry and particle identification
in each section: the first stage detects low momentum
particles with large angles ($\pm180$ mrad).  High momentum particles
are analyzed in the second part ($\pm$25 mrad) using a large lever arm
and a higher magnetic field than in the first stage. In this way an
angular resolution down to 10~$\mu$rad (for small scattering angles)
and a transverse resolution down to 7 $\mu$m can be achieved.  A schematic
view of the spectrometer is given in fig. \ref{fig:setup}. For the
double charm setup the gap of the first spectrometer magnet should
be reduced from the present 1.72 m to 0.82 m, which is possible by
removing some of the modular yoke pieces. This provides a higher field
for the higher beam energy and also a smaller stray field in the
tracking zones.

The tracking system is built up from a set of {\bf L}arge {\bf A}ngle
{\bf T}rackers ({\bf LAT}) covering the outer region with lowest track
density and {\bf S}mall {\bf A}ngle {\bf T}rackers ({\bf SAT}) for the
inner regions.  So-called {\em tracking stations} are distributed all
over the spectrometer and consist of three different detector types
staggered, each smaller one having finer granularity and rate
capability and covering with some overlap a central hole in the next
larger one.  For the innermost part, directly in the beam, silicon
detectors \cite{elmau} and scintillating fibers are used. The LAT are
small cell size drift chambers, straw chambers\cite{straws} and
further downstream MWPC and large cell size drift chambers. The very
important inner trackers between the beam region and the LAT consist
of Micromegas \cite{micromegas} before the first magnet and GEM detectors
\cite{gem} after that.

The compatibility of the tracking system with the high intensity
hadron beam, however, still has to be proven. In particular the inner
trackers may run at a higher risk of discharges, which could make
a partial revision of the setup necessary.

Particle identification is first of all given by RICH 1. A second RICH
is currently under planning and should be able to cover the high
momentum part of the spectrum passing through the second magnet. It is
foreseen to have a fast readout which could make particle
identification at the second trigger level possible. In addition
two upgraded Cherenkov counters of CEDAR type will provide particle
identification in the beamline to be able to obtain clean cross section
measurements.

Further downstream electromagnetic and hadronic calorimeters provide
the energy measurement needed to form a first level trigger based on
transverse energy. Thereafter two muon filters allows to identify muons
from semi-leptonic charm decays. In addition to the existing muon
walls the first muon filter has to be augmented by a muon hodoscope
for triggering. Only a small fraction of muons from semi-leptonic
decays would reach the present muon hodoscopes put far downstream for
DIS measurements.
\subsection{Target Setup}
\begin{figure}[h]
  \begin{tabular}{cc}
    \includegraphics[angle=90,width=7cm]{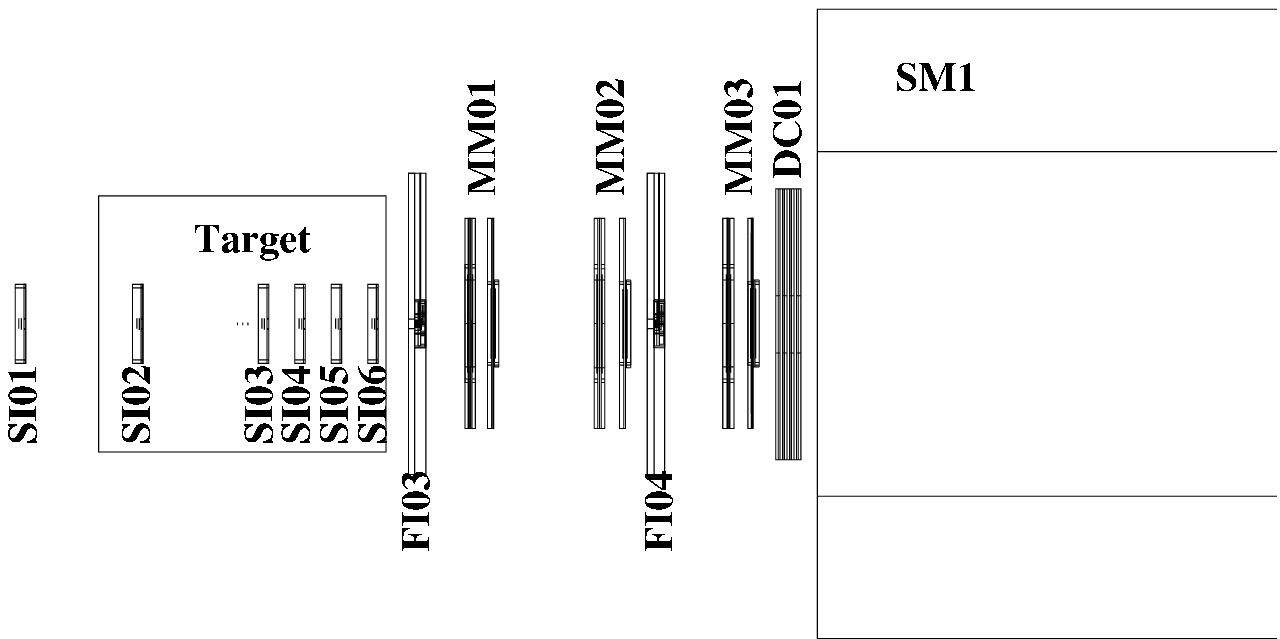}&
    \includegraphics[width=7cm]{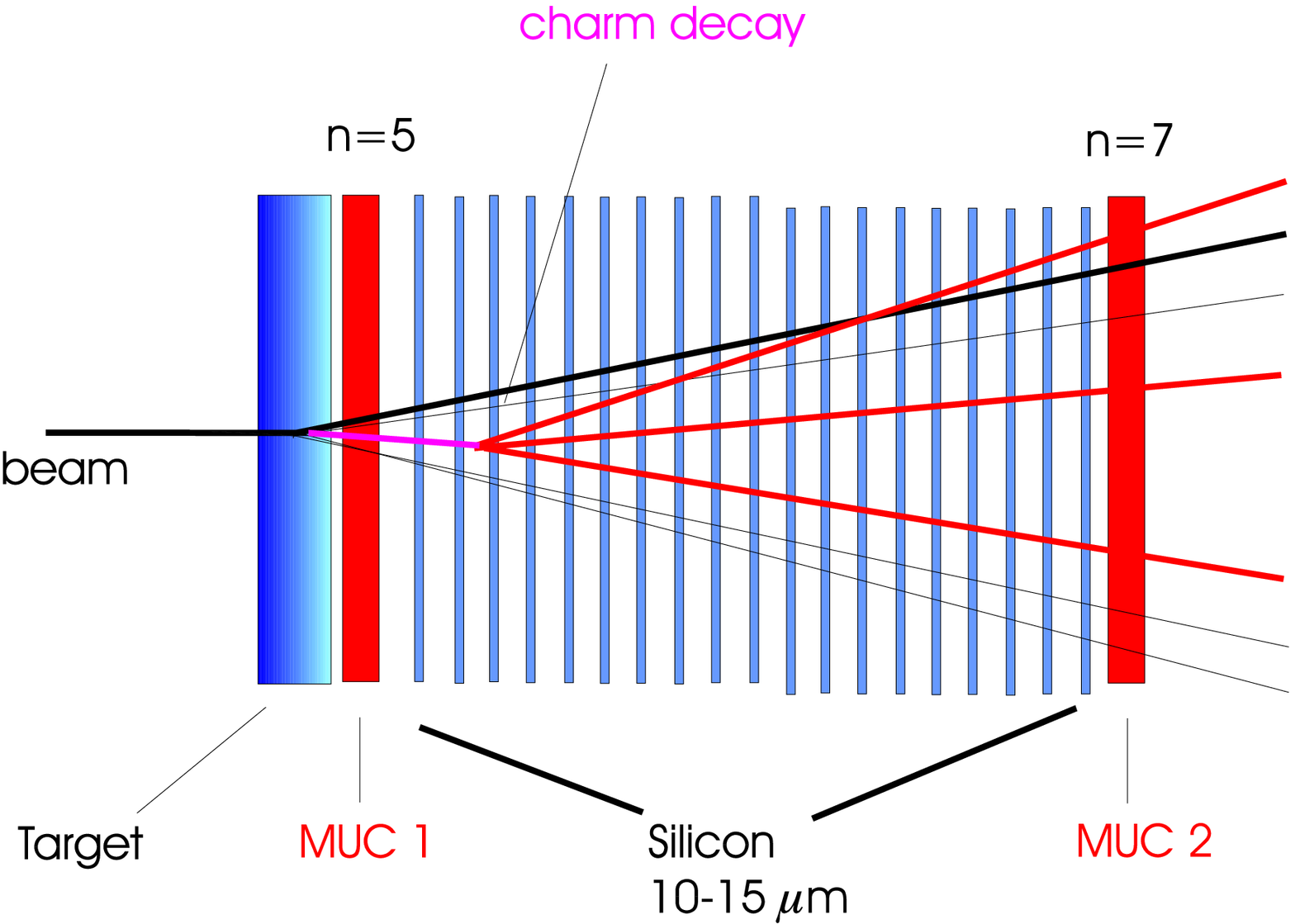}\\
    \begin{minipage}[t]{7.5cm}
      \caption{Target region as used in the simulation.}
      \label{fig:target}
    \end{minipage}&
    \begin{minipage}[t]{7.5cm}
      \caption{Charm decay detector. 16-20 planes are spaced by 2 mm and
        have a pitch of 10 -- 15 $\mu$m.}
      \label{fig:decaydet}
    \end{minipage}
  \end{tabular}
\end{figure}

The target setup as described in the present simulation is shown in
figure \ref{fig:target}. Beam definition is provided as in the present
setup by scintillating fibres and silicon microstrips.  The beam
impinges on a segmented target of in total 2\% of a nuclear
interaction length.  It is foreseen to use different materials for the
thin target plates to study the A-dependence of charm production.
The segmentation allows charm hadrons to decay mostly outside the target
material to allow a cleaner vertex separation.

After the target a silicon microstrip telescope is needed to allow
precise vertex determination. A clear separation of primary production
vertex and secondary decay vertex is the cleanest signature for a
weak charm decay. As a minimum setup 16 microstrip planes arranged in
four projections are foreseen. Depending on resolution, acceptance and
possible stray fields a fifth station might be needed. A further
consideration is the possibility to fully reconstruct semi-leptonic
decays by using a densely packed decay detector right after the target
as shown in figure \ref{fig:decaydet}. In the order of 16 planes are spaced 
within 2 cm along the beam and have a pitch of 10 -- 15 $\mu$m. This
setup would allow to see a fraction of charm tracks (mostly $D$-mesons) 
still in the detector before their decay.
Both resolution and the benefit of a charm decay detector still have
to be studied carefully in simulations. 

After the target a scintillating fibre detector will be used to
obtain the track multiplicity at the trigger level and provide 
precise timing for all vertex tracks.

In COMPASS no precision vertex detector is present yet. The design of
this detector has to fulfill a number of strict requirements:
\begin{itemize}
\item The detector has to stand fluences up to $5\times 10^{14}$
  particles/cm$^2$.
\item The spatial resolution in beam direction should be better than
  100 $\mu$m to provide sufficient resolving power for charm decay
  vertices. 
\item Finally, at the very high particle rates foreseen a good timing
  resolution is needed to recognize interaction pileup and disentangle
  multiple beam tracks.
\end{itemize}
It is foreseen to make use of the Lazarus effect \cite{lazarus} by
operating silicon detectors at cryogenic temperatures so that they
can survive larger fluences.

During the process of detector optimization we have to investigate
the design of a monolithic target-vertex-cryostat, determine the
best pitch size and determine the effects of a larger lever arm
vs. acceptance and mechanical design.

Finally the readout has to run at a speed of up to 100 kHz. ADCs
are needed to obtain better space resolution, a good timing of the 
signal and discrimination of secondary interactions.

\section{TRIGGER SCENARIO}
One of the most important problems to solve is to reduce the vast
number of inelastic interactions to the interesting ones showing
signatures for charm hadron decays.  This is discussed in this 
section.

The starting point are $2\times 10^6$ interactions per SPS spill.
The first level trigger has to reduce this to a rate of not more
than 100 kHz, i.e. by at least a factor 4. Due to the constraints
of the readout system of existing detectors per event 1 $\mu$s is
available for the trigger decision at this level.

Three types of triggers can be envisaged at this trigger level, of which
a combination should be able to reach the desired rejection level at
a reasonable efficiency:
\begin{itemize}
\item Requiring simple charged track multiplicities bigger than 4 is
a safe cut, in particular when searching for double charm. This removes
a large part of the diffractive inelastic reactions and all elastic ones.
\item A large transverse energy detectable by the calorimeters can
  indicate charm decays.
\item A rather high fraction (up to 17\%) of $D$-mesons decay
  semi-leptonically, mostly producing a muon. This can be selected by
  the muon filters of the spectrometer.
\end{itemize}
These trigger types shall be discussed in the following.

The rate coming from the first level trigger is still far too high
to be written to mass storage. Therefore a second and/or third trigger
level is required. The features exploited at these further levels are
described below:
\begin{itemize}
\item Hit multiplicity, suppression of secondary interactions and a
  possible multiplicity jump in the vertex detector can be already
  performed in an intelligent detector frontend.
\item Track prototypes can be formed within a super ROB, a
  particularly powerful readout buffer computer which reads the entire
  vertex detector. With these track angles, track multiplicity and
  possibly high track impacts can be investigated.
\item Vertex reconstruction needs the power of a third level trigger
  farm with many CPUs or high power co-processor cards for the
  super ROB. Here vertices can be searched and separated
  production and decay vertices can be identified.
\item Finally particle identification can be performed from the
  information of the RICH detectors and secondary vertices of hyperons
  and strange mesons can be tagged.
\end{itemize}
The final rate should be in the order of 10--20,000 triggers per spill.
\subsection{Transverse Energy}
\begin{figure}[h]
  \begin{tabular}{cc}
    \includegraphics[width=7cm]{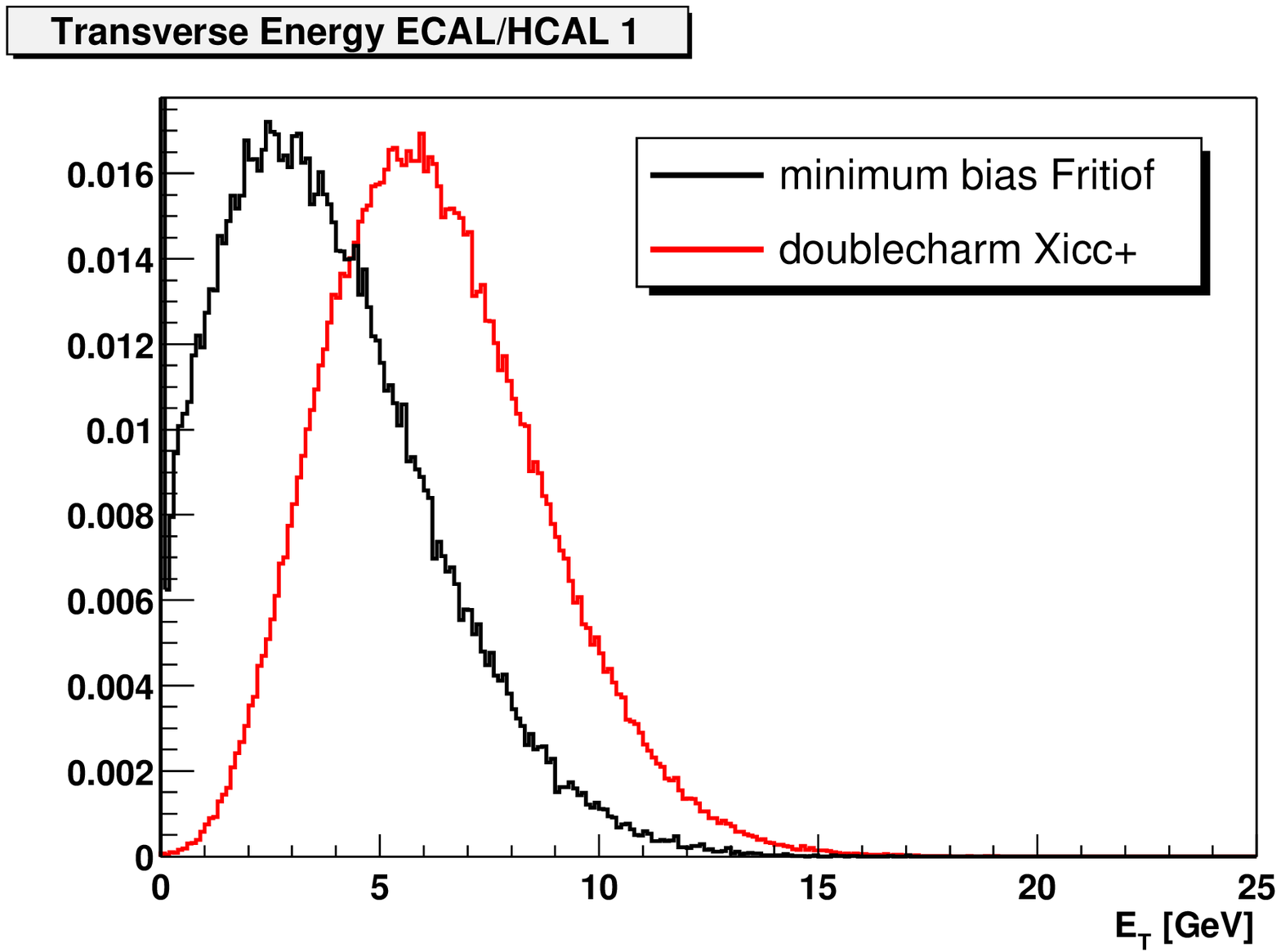}&
    \includegraphics[width=7cm]{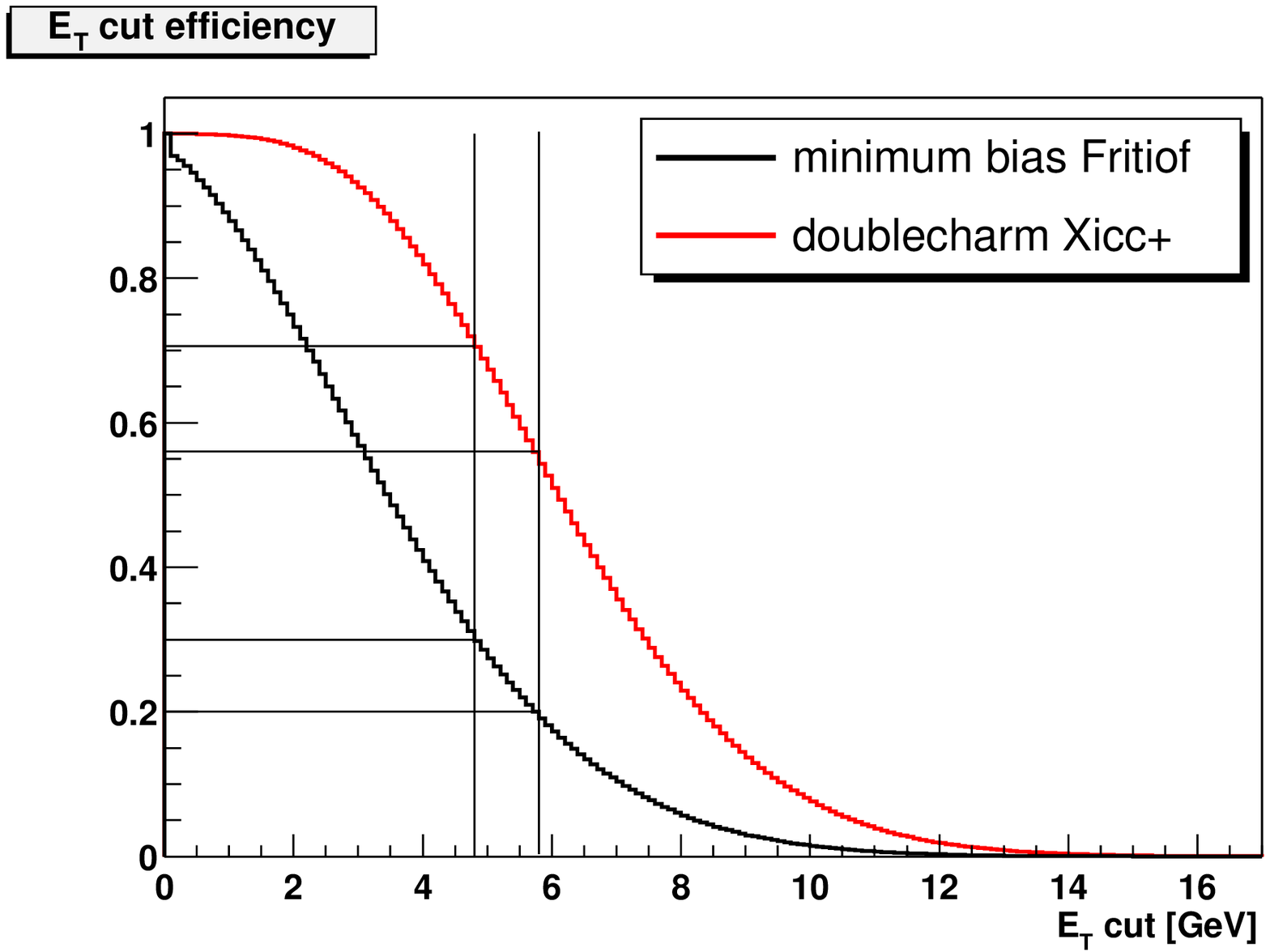}\\
    \begin{minipage}[t]{7.5cm}
      \caption{Distribution of transverse energy of minimum bias
        events and double charm events.}
      \label{fig:et}
    \end{minipage}&
    \begin{minipage}[t]{7.5cm}
      \caption{$E_T$-cut efficiency. Shown are curves for minimum bias
        and double charm. Intersections for a background reduction to
        20\% and 30\%  are drawn.}
      \label{fig:eteff}
    \end{minipage}
  \end{tabular}
\end{figure}

The high charm quark mass of around 1.5 GeV opens up a large number of
decay channels. At the same time the Q-value of the decay is large and
therefore also the transverse momenta of the decay products. Enriching
events with high $p_T$ tracks will therefore also enhance charm decays.
Experiments E791 and E831 already used this type of cut successfully.
Typical values are 3 to 5 GeV reducing the number of triggers by a
factor of 3 to 5 at efficiencies between 70 and 55 \%. The simulation
of doubly charmed baryons shows an even higher transverse energy due
to the two charm quarks and rather long decay chains. Figures \ref{fig:et}
and \ref{fig:eteff} illustrate this trigger.

\subsection{Multiplicities and Muon Trigger}
\begin{figure}[h]
  \begin{tabular}{cc}
    \includegraphics[width=7cm]{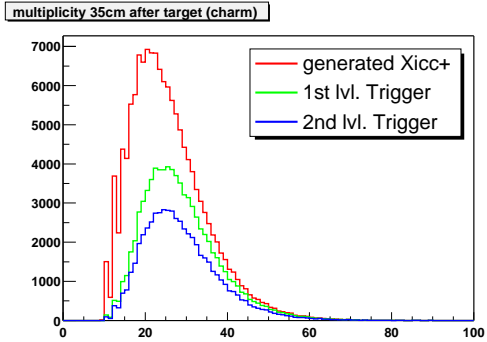}&
    \includegraphics[width=7cm]{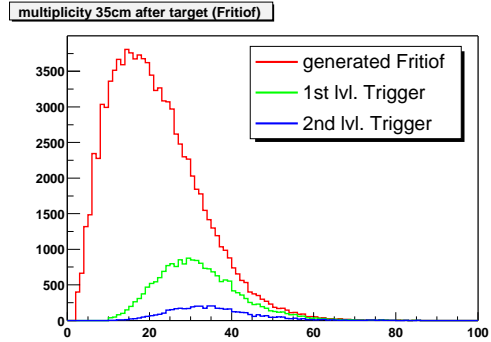}\\
    {\footnotesize a) Multiplicities from double charm} &
    {\footnotesize b) Multiplicities from minimum bias}\\
  \end{tabular}
  \caption{Multiplicities for all generated events, events passing a
    first level trigger consisting of a multiplicity cut plus a
    minimum transverse energy and events requiring a reconstructible
    secondary vertex are compared.}
  \label{fig:multi}
\end{figure}

Due to the large Q-value of the double charm decay and the long decay
chains the charged track multiplicity is very high. For example
looking at the decay $\Xi_{cc}^+\rightarrow\Lambda_c K^-\pi^+$ with
some associated $D$-mesons basically no events with less than 10 charged 
tracks are seen. A simple cut of at least 4 tracks would already cut
down the events by half with 100\% charm efficiency. Higher reduction
factors at still very high double charm efficiencies can be reached.
Technically this trigger would be implemented by means of a fast
scintillating fibre detector at the end of the vertex
region. Multiplicities of double charm events and minimum bias events
are compared in figures \ref{fig:multi} a) and b).

Another clean signature is the production of muons in semi-leptonic
decays of charm mesons. The branching ratios are
\[
BR(D^0\rightarrow\mu X) = 7\%\ ,\quad\text{and}\quad
BR(D^+\rightarrow\mu X) = 10\%\
\]
whereas background from pion and kaon decays is small due to their
much longer lifetimes compared to the charm mesons. Therefore a
reduction factor of 30 from $\sigma_{tot}$ can be reached at an efficiency
which is basically equivalent to the semi-leptonic branching ratio.
In addition background can be further reduced by requiring a minimum
transverse momentum of the muon.

\subsection{Online Filter}
Coming to an acceptable data rate requires online filtering of the
events.  In the present COMPASS DAQ system with its 12 (to max. 16)
eventbuilder computers (with 2 processors each) there is little room for
complicated tasks, since at 40 kHz only 2-3 ms are available to
process one event.\\
Therefore other, additional or alternative ways of filtering and data
processing have to be implemented. The various possible options shall
be briefly discussed here.

A very resource efficient first approach is to improve the frontend
electronics of detectors relevant to further trigger decisions with
more processing power by means of fast Field Programmable Gate Array
(FPGA) chips. This allows the preprocessing of data at an early stage
saving CPU power on the actual filtering stage for mostly physics
oriented data treatment. Possible tasks for frontend preprocessing are:
\begin{itemize}
\item Correlation and cut on the signal time;
\item Data reduction by forming clusters from adjacent channels
  including time cuts and amplitude weighting;
\item Determination of hit multiplicities;
\item Rejection of secondary interactions by means of a second
  threshold for too large signals.
\end{itemize}

As a second level trigger a special readout buffer computer (Super {\em ROB})
could be developed for the vertex detector. A large part of physically
relevant data arrives at this single computer and equipping it with
extra CPU power (4--8 CPUs) and moreover powerful DSP or FPGA
co-processor cards can yield a high selectivity based on simple criteria.
Mainly the forming of track prototypes can be performed here from
which selections on
\begin{itemize}
\item track multiplicities,
\item track angles (partly correlated to transverse momenta),
\item a preliminary interaction vertex and
\item high track impact parameters
\end{itemize}
can be derived easily. This can be nicely embedded in the COMPASS DAQ
system by also making this special machine the event distribution manager
which directs the data flow from ROB computers to eventbuilders. A large
fraction of events would be simply flagged by the EDM after processing 
on the Super ROB to be discarded on all other ROBs.

The final filtering of events can be done in a dedicated filter farm
consisting of densely packed CPUs, either as flat rack servers or 
better server blades with CPUs with low power consumption in racks
with high integration and built-in network, power and cooling 
infrastructure. These systems can attack complicated filtering tasks 
like
\begin{itemize}
\item Secondary vertex reconstruction;
\item Tagging of daughter decays (hyperons, $K^0_s$, $D$-mesons);
\item Reconstruction of RICH rings.
\end{itemize}
In a system with 300--600 CPUs processing time of up to 25--50 ms per
event is available.

\begin{figure}[h]
\begin{center}
\includegraphics[width=10cm]{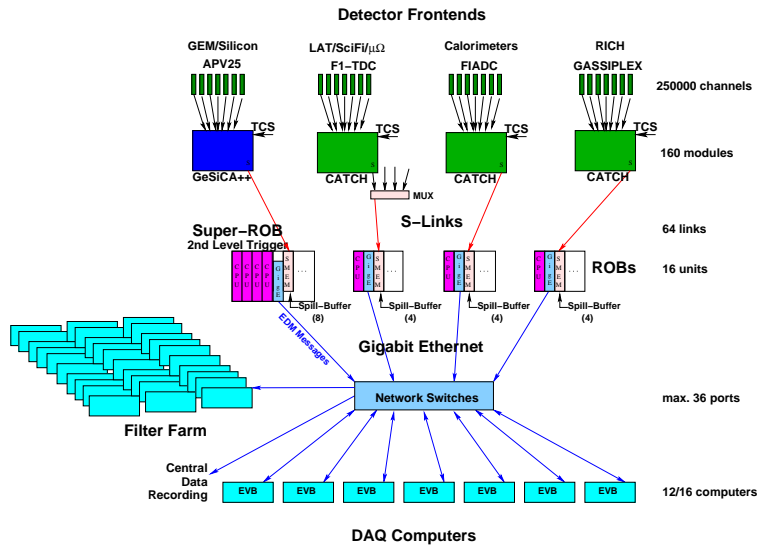}
\caption{Improved DAQ system for the second phase of COMPASS.}
\label{fig:daq}
\end{center}
\end{figure}

An alternative approach can be the implementation of a more
homogeneous system of networked compute nodes which in a first layer
address directly buffered detector data and then pass on data to
further levels \cite{s-daq}.  This approach in its full reach can even
accomplish a readout system without dedicated hardware trigger
signals, which samples data at a constant frequency and performs data
reduction, feature extraction and filtering in parallel as the data is
transported and combined.  This however puts also strong requirements
on the frontends which actively have to perform hit-detection and data
reduction already before transferring any data. Although this might not
be fully realizable in COMPASS the compute node network can be a cost
effective alternative to an expensive CPU farm.

\section{SIMULATIONS}
In preparation for the COMPASS Future Workshop in September 2002
a number of simulation studies have been performed. Their results
are summarized here. Further studies to address a number of open
questions are under way.\\

\begin{figure}[h]
\begin{center}
\includegraphics[width=10cm]{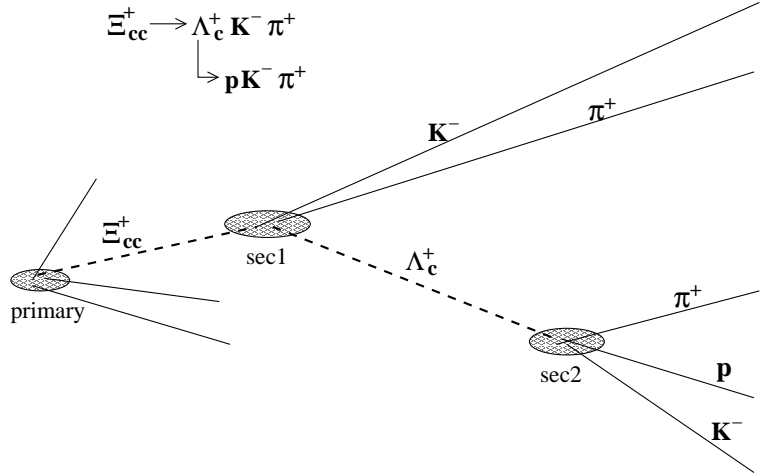}
\caption{Schematic decay chain of the $\Xi_{cc}^+$ in the SELEX channel.}
\label{fig:xicc}
\end{center}
\end{figure}

The first set of simulations of doubly charmed baryons in COMPASS was
based on the decay channel $\Xi_{cc}^+\rightarrow\Lambda_c K^-\pi^+$
observed by SELEX at FNAL \cite{mattson}. Together with the baryon two
anti-$D-$mesons are generated. The production parameters are assumed
to follow
\begin{align*}
\sigma &\sim (1-x_F)^3\\
\sigma &\sim \exp(-1.2p_T^2)\quad .
\end{align*}
The remaining energy is given to the Fritiof event generator to add
further light hadrons.

For the detector simulation {\em COMGEANT}, a Monte Carlo program
based on GEANT 3.14 is used. The detector geometry used here is from
the first spectrometer magnet SM1 onwards identical to the present DIS
setup.  The target area before this magnet was already shown in figure
\ref{fig:target} but there is further room for optimization.  Currently
only fieldmaps for magnet gaps of 1.72 m and 1.32 m are available.
However a more favorable gap of 82 cm should be studied in the near
future.

\noindent The following assumptions and cuts are implied in this 
simulation study.
\begin{itemize}
\item The doubly charmed baryon was simulated with a lifetime of 25 fs
  corresponding to a value as favored by the SELEX observations.
\item The main cuts in GEANT were set to 100 MeV for faster processing.
\item There was no detailed simulation of the RICH, only momentum
  thresholds for the various particles were applied. Positive
  identification of all kaons and protons was required.
\item Any secondary charm vertex was required to be outside the target
  material to be reconstructible.
\end{itemize}
The following sections illustrate the results of these simulations.
\subsection{Acceptance and Resolution}
The overall geometrical acceptance and tracking capability of the
simulated setup was found to be 5\%. Detector efficiencies were not
yet applied, but the present reconstruction program CORAL was used.

\begin{figure}[h]
  \begin{tabular}{cc}
    \includegraphics[width=7cm]{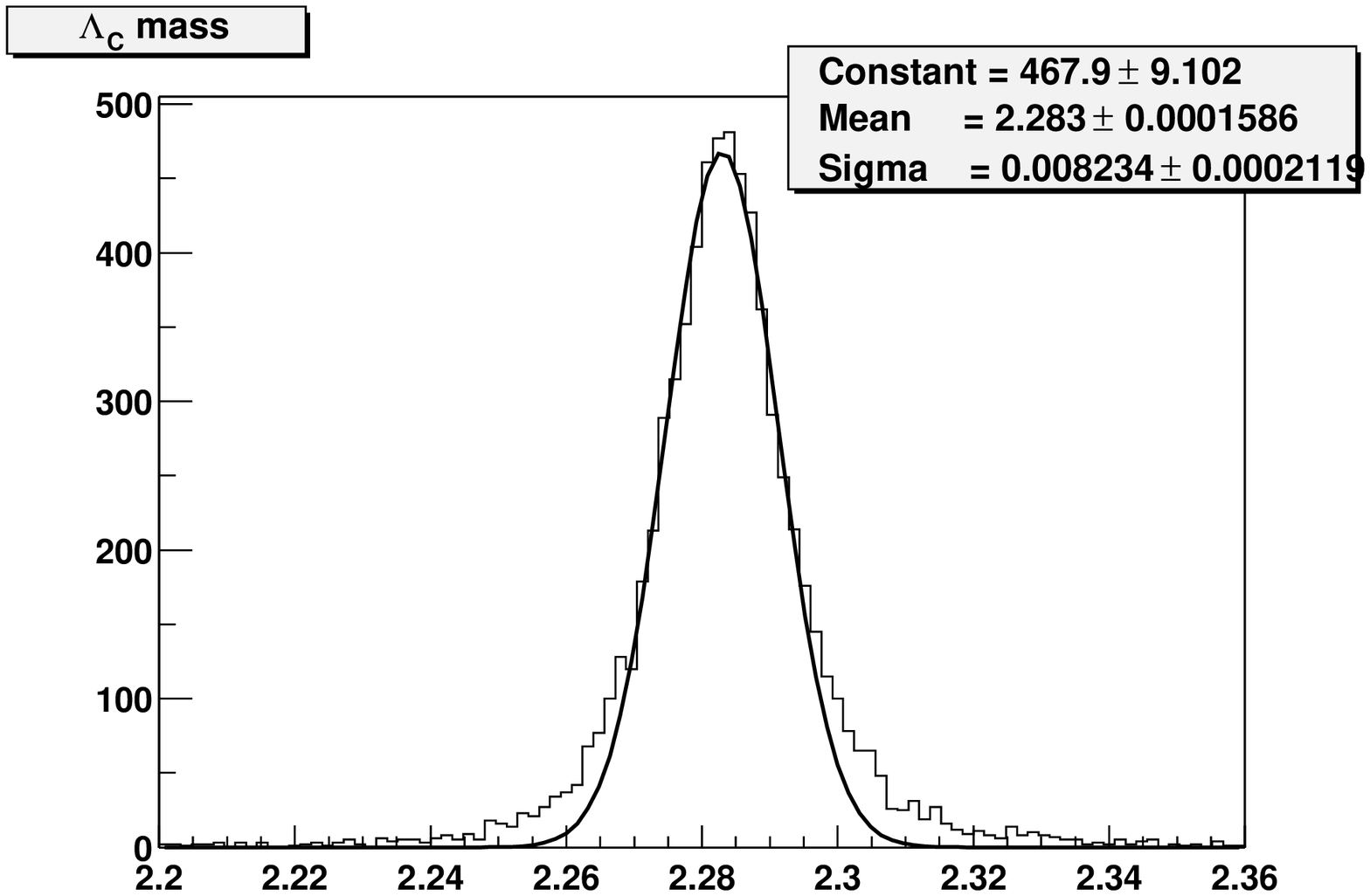}&
    \includegraphics[width=7cm]{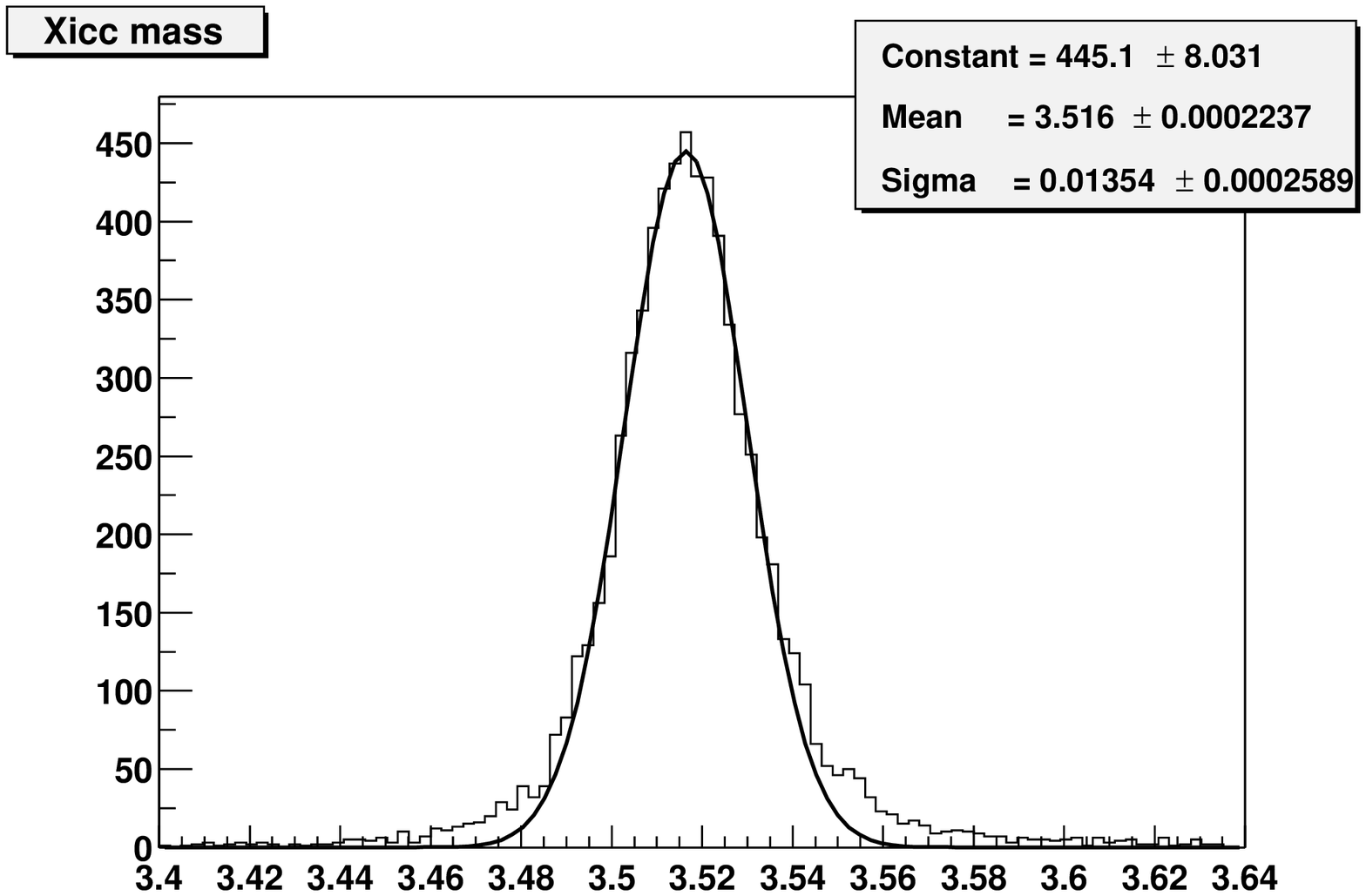}\\
    {\footnotesize a) $\sigma(\Lambda_c)=8$ MeV}&
    {\footnotesize b) $\sigma(\Xi_{cc})=13$ MeV}\\
  \end{tabular}
  \caption{Mass resolutions of $\Lambda_c$ and $\Xi_cc$ after
    reconstruction.}
  \label{fig:massres}
\end{figure}

The mass resolution for the reconstructed $\Lambda_c$ was found to be
8 MeV whereas the resolution of $\Xi_{cc}$ turned out to be 13 MeV as
shown in figure \ref{fig:massres}.

\subsection{Momenta and Track Efficiencies}
\begin{figure}[h]
  \begin{tabular}{cc}
    \includegraphics[width=7cm]{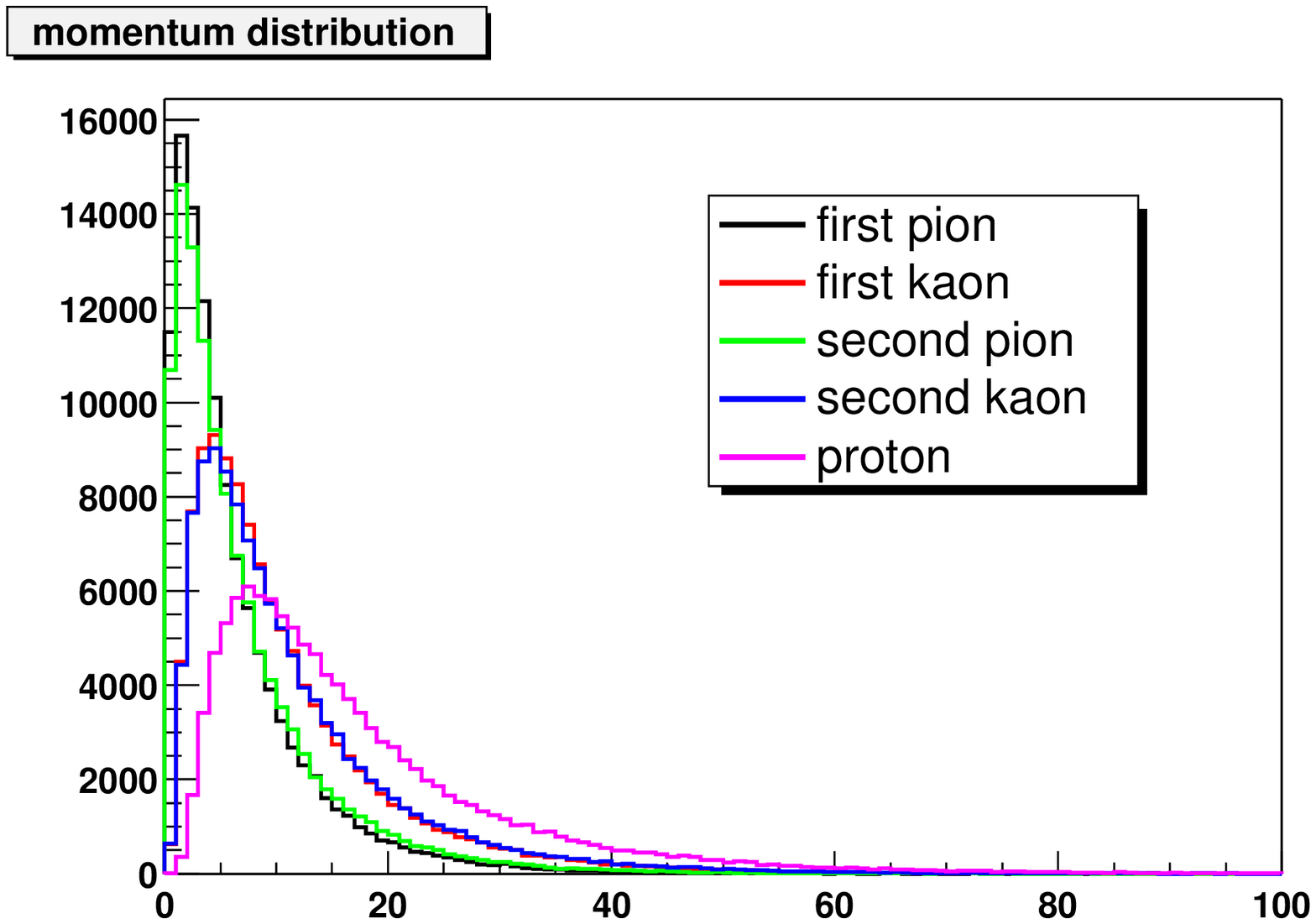}&
    \includegraphics[width=7cm]{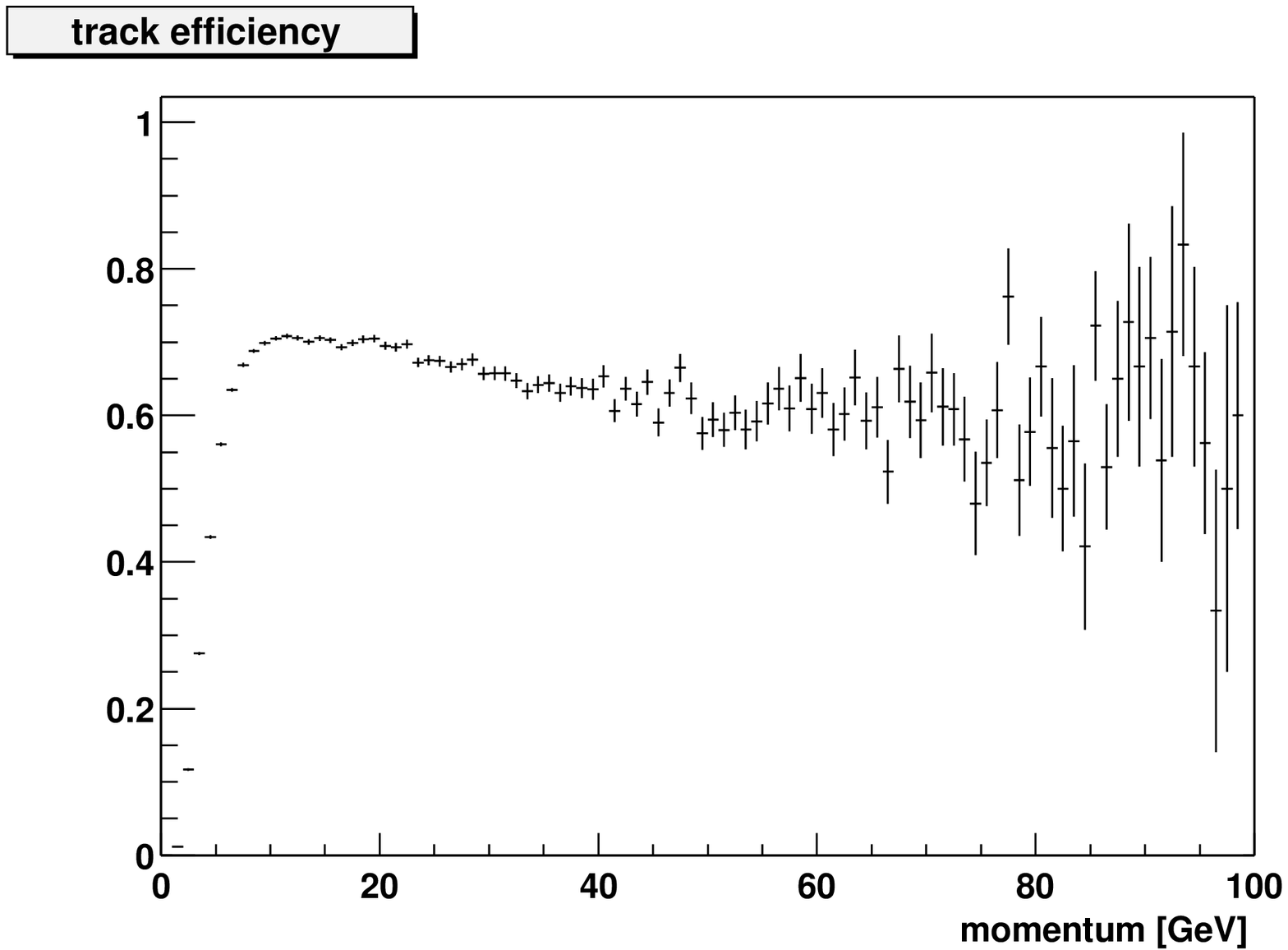}\\
    \begin{minipage}[t]{7.5cm}
      \caption{Distribution of momenta of particles arising form the
        $\Xi_{cc}$ decay.}
      \label{fig:momenta}
    \end{minipage}&
    \begin{minipage}[t]{7.5cm}
      \caption{Tracking efficiency vs. momentum for the present setup.}
      \label{fig:trackeff}
    \end{minipage}
  \end{tabular}
\end{figure}

Figure \ref{fig:momenta} summarizes the distribution of momenta of the
various particles simulated in the $\Xi_{cc}$ events. It is notable,
that only very few particles reach momenta above 40 GeV/c. This
underlines the importance of the first spectrometer magnet. Presently a
relatively poor average momentum resolution of 2\% in this region is found 
indicating that a reduced gap of SM1 leading to a higher field would be 
beneficial. Figure \ref{fig:trackeff} shows that for particles above 5 GeV/c
a reasonable tracking efficiency above 80\% can be reached with the
present setup.

\subsection{Trigger Efficiencies}
Finally it was also possible to derive in a simplified way estimates
for trigger efficiencies from the simulations. Note however, that in
the following exclusive double charm events are only compared to events
generated by Fritiof as a kind of minimum bias hadronic background.
A large part of the total cross section constituted by elastic and
diffractive scattering events are not treated here. They should be
strongly suppressed by a hard multiplicity cut.

\begin{center}
\begin{tabular}{l|r}
{\bf Trigger type} & {\bf Ratio Charm/Fritiof}\\
\hline
{\bf Muon trigger} ($p(\mu)>2$ GeV/c) & 29.5\% / 11.7\%\\
{\bf Multiplicity trigger} & 100\% / 85\%\\
\quad(more than 10 charged tracks) & \\
{\bf 1st level trigger} & 57.7\% / 20 \%\\
\quad({\it Multiplicity} 
$\wedge\ ((E_T > 5.8 \text{GeV}) \vee (E_T > 3 \text{GeV} \wedge
\mu))$) & \\
{\bf 2nd level trigger} & 41.5\% / 4.6\%\\
\quad(some vertex activity) & \\
\end{tabular}
\end{center}
These values show that already in comparison to the relatively hard
spectrum of products from Fritiof the discussed trigger scenario would
work, i.e. provide a sufficient reduction at the first trigger level 
and good selectivity at the second.

\begin{figure}[h]
  \begin{tabular}{cc}
    \includegraphics[width=7cm]{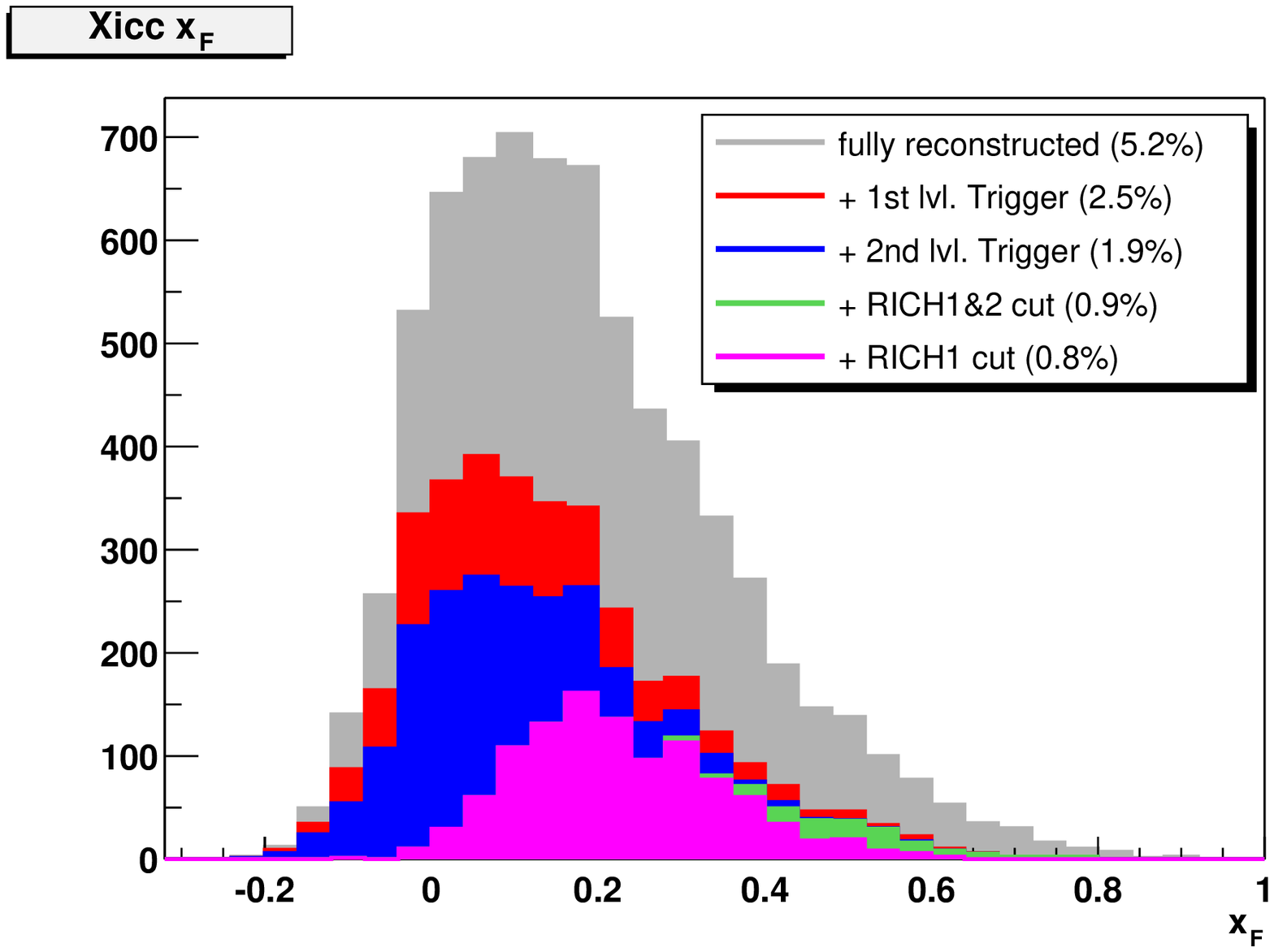}&
    \includegraphics[width=7cm]{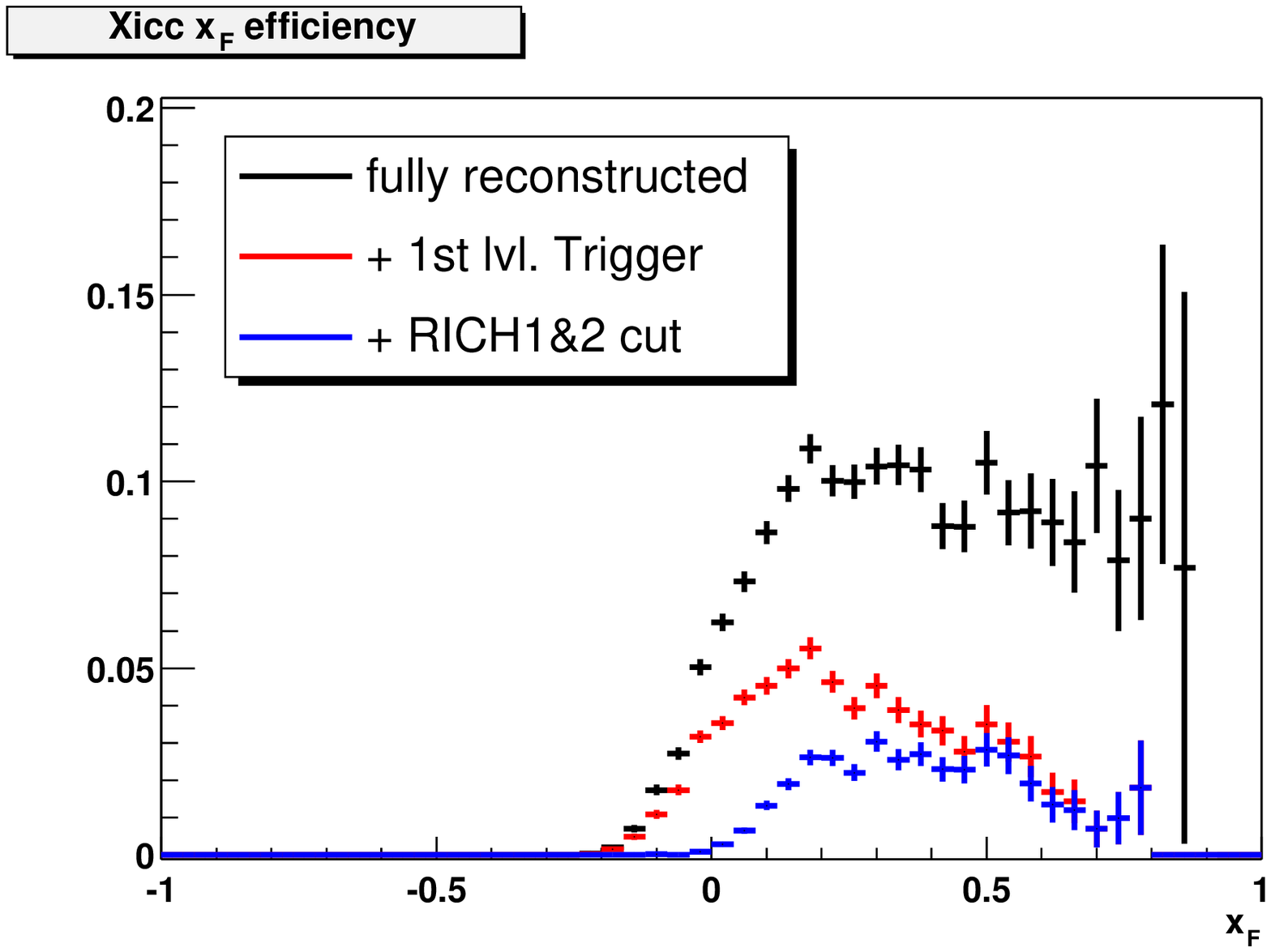}\\
    \begin{minipage}[t]{7.5cm}
      \caption{Accepted events vs. $x_F$ for the various steps in the
    triggering process. In addition the effect of RICH cuts based on 
    simple acceptance and momentum thresholds are shown.}
      \label{fig:xiccxf}
    \end{minipage}&
    \begin{minipage}[t]{7.5cm}
      \caption{Reconstruction and trigger efficiencies normalized to
    all simulated events.}
      \label{fig:xiccxfeff}
    \end{minipage}
  \end{tabular}
\end{figure}

In addition figure \ref{fig:xiccxf} shows the absolute
$x_F$-acceptance of reconstruction and triggers and figure \ref{fig:xiccxfeff}
the efficiency distribution normalized to all generated events. Here
also the effect of RICH cuts was included. These cuts are based simply
on acceptance and momentum cuts according to the operating thresholds
of the employed RICH detectors. It turns out that the second RICH
detector does not contribute a lot, mostly due to the fact, that
the fraction of tracks with momenta above 40 GeV/c is rather low. 
Nevertheless the high momentum part is of particular interest for the
highly forward produced subsample which is of biggest relevance in the
comparison with the SELEX results. In any case RICH cuts in the end would 
clean up the data substantially and reduce ambiguous interpretations.

The figures show as well that a reasonable reconstruction efficiency
is reached already at $x_F>-0.1$.

\section{RATE ESTIMATES}
From the conducted simulations rate estimates can be obtained. Further
input are the nominal beam rate of up to $10^8$ protons per spill and
the assumed target with a thickness of 2\% of an interaction length.
This leads to a total of $10^{12}$ interactions in a run of 100
effective days which corresponds to an integrated luminosity of
$\int{\cal L} = 25$ pb$^{-1}$, based on a total cross section of 
40 mb per nucleon.

The SELEX observations (\cite{russ,cooper}) point to an unexpectedly
large fraction of double charm production: From roughly 1600
reconstructed $\Lambda_c$ they obtain about 50 doubly charmed baryons.
If one then takes into account all cuts and branching ratios of the
observed channels one comes to the conclusion that about half of all
$\Lambda_c$ in fact come from doubly charmed baryons. Assuming
therefore a double charm production cross section in the order of the
singly charmed baryon production cross section of about 2 $\mu$b (c.f.
results from WA89 \cite{wa89} at a similar energy as COMPASS) this would 
mean for COMPASS, that 50 million doubly charmed baryons would be 
produced.

Taking now into account the results of the simulation (acceptance
$\times$ reconstruction $\times$ trigger $\times$ RICH= 0.8\% as from
figure \ref{fig:xiccxf}) and estimates for
branching ratios ($BR(CCQ)\times BR(CQQ)=30\%\times 20\%=6\%$) and
assuming an additional factor for the overall detection and vertexing
efficiency of 40 to 70\% one arrives at 10000 to 17000 reconstructed
doubly charmed baryons. Here $BR(CCQ)$ already includes an estimated
sum of all measurable CCQ channels, not only the simulated channel
$\Xi_{cc}^+\rightarrow\Lambda_c K^-\pi^+$ and $BR(CQQ)$ denotes the
fraction of reconstructible daughter baryon decays.  This result is quite
remarkable aside of the simple numbers in the sense that this would
bring the highly interesting field of $CCQ$ spectroscopy into reach.

But even a more conservative cross section estimate along the lines of
$\sigma(CCQ) \sim \sigma_{tot}\times(10^{-3})^2$, i.e. a factor
$10^{-3}$ down from charm production giving in the order of 10 nb
would correspond still to 250000 produced or 100 to 170
reconstructed doubly charmed baryons. This number nevertheless would
constitute a solid observation.

\section{CONCLUSIONS}
In this report the hardware requirements for the measurement of doubly
charmed baryons in COMPASS were outlined and the results of first
simulations were presented. It was shown, that sufficient suppression
factors for the first trigger level could be reached by a combination
of multiplicity and transverse energy cuts enhanced by identifying
muons from semi-leptonic decays. Based on these results rate estimates
for a COMPASS measurement were obtained.
With the exiting SELEX observations in view doubly charmed baryons could be
produced so abundantly, that $CCQ$-spectroscopy would be in reach
for COMPASS.

Simultaneously to the search for doubly charmed baryons valuable
high statistics data on singly charm baryons can be obtained. Here
one has to choose between a single charm measurement with double 
charm as bonus or a strict orientation of setup and triggers toward 
double charm. This choice is essentially determined by data rates and 
the desired selectivity.

However there is still a lot of work to be done. An optimized setup
has to be found for the vertex detector, the spectrometer layout and
the trigger detectors. In particular there is substantial design work
needed for the vertex detector. Further simulation studies
of trigger efficiencies and reduction factors are needed. It would be
interesting to obtain a better handle on the rejection of minimum bias
events from real hadron beam data. Then the best filter algorithm has
to be found and coded. Further question marks lie in lifetimes, cross
sections and production mechanisms of doubly charmed baryons.

\end{document}